\documentstyle[preprint,revtex]{aps}
\tightenlines

\begin{document}
\draft
\preprint{
\parbox[t]{2.5in}{
SU-ITP-9233, astro-ph/9304013 \\
Submitted to Physical Review D \\ {}
}
}

\begin{title}
A Green Function for Metric Perturbations \\
due to Cosmological Density Fluctuations
\end{title}

\author{
Mark W. Jacobs,\cite{mwj}
Eric V. Linder,\cite{evl} \\
and Robert V. Wagoner\cite{rvw}
}

\begin{instit}
Department of Physics and
Center for Space Science and Astrophysics, \\
Stanford University, Stanford CA 94305
\end{instit}

\receipt{4-6-93}

\begin{abstract}
We study scalar perturbations to a Robertson-Walker cosmological
metric in terms of a pseudo-Newtonian potential, which
emerges naturally from the solution of the field equations.
This potential is given in terms of a Green function for matter
density fluctuations of arbitrary amplitude whose time and
spatial dependence are assumed known.  The results obtained
span both the linearized and Newtonian limits, and do not
explicitly depend on any kind of averaging procedure, but make
the valid assumption that the global expansion rate is that of
a Friedmann-Robertson-Walker model.  In addition, we discuss
the similarity to diffusive processes in the evolution of the
potential, and possible applications.
\end{abstract}
\pacs{PACS number(s): 98.90.Dr,04.20.Jb}

\narrowtext
\section{Introduction}
Most observers would agree that the geometry of our universe is
well-described by perturbations to a Robertson-Walker metric.
These perturbations directly affect the information we receive
{}from the universe outside of our own galaxy, and so a great
deal of effort has been devoted to understanding their effects
and evolution.

Here we study the effects of a given matter distribution on the
metric, and hence on the radiation that reaches us.  Our goal
is to find a way of expressing these perturbations directly
in terms of the matter variables, in a way that may be useful
for interpreting or modeling observations.  Most simulations
of observable effects---gravitational lensing, redshift, and so
on---use some kind of radically simplifying assumption, such as
nearest-neighbor Newtonian gravitation, weak density contrast
(i.e.\ linearized theory), etc.  Here we attempt to provide an
expression which takes account of all cosmologically relevant
effects, and which applies over a wide range of density contrasts.

This paper expands on the results presented briefly in reference
\cite{jlw92}.  We show how to derive the Green function presented
there for any value of the curvature parameter, as well as its
reduction to the Newtonian limit and the similarity to simple
diffusion.  The principal results are summarized by equations
(\ref{eq:metric2}),
(\ref{eq:phiG}),
(\ref{eq:genG}),
and (\ref{eq:G0})
for the perturbed line element, pseudo-Newtonian potential,
and Green function.

\newpage
\section{Conventions and Definitions}
Units are chosen with $G=c=1$ so that all quantities can be
expressed as powers of a length.   The metric is written as
a perturbation on the static part of the Robertson-Walker
background, $\gamma_{ab}$, with signature $+2$ and curvature
parameter $k\in\{-1,0,1\}$:
\begin{equation}
 ds^2 = a^2(\eta)
 [\gamma_{ab}(\vec{x}) + h_{ab}(\eta,\vec{x})] dx^a dx^b\ .
\label{eq:metric}
\end{equation}
The coefficients and coordinates are dimensionless, so the scale
factor $a(\eta)$ (which we write as a function of conformal
time) is a length.  Indices $a,b,c,\ldots$ run 0--3, except for
$i,j,\ldots,n$ (the Fortran integers) which range over the spatial
indices 1--3.  An extremely useful source of formulas for the
background metric and perturbations, in both conformal and proper
time, is appendices A-D of Kodama and Sasaki, 1984~\cite{ks84}.
Our notation is similar; in particular a prime (e.g.\ $a'$) always
means derivative with respect to conformal time.

Naturally the perturbations $h_{ab}$ are assumed small compared
to $\gamma_{ab}$, which are of order unity.  Of course this does
not imply that the corresponding matter density fluctuations are
small compared to the background.  We use the scheme of Futamase
1988~\cite{fut88} for parameterizing the size of the perturbations
and their derivatives, which is very much like the one used in
the shortwave approximation for studying gravitational waves
in a curved background (see Isaacson, 1968a~\cite{isac68a}).
With this scheme, orders of magnitude in our problem are:
\begin{mathletters}
\begin{eqnarray}
 \gamma_{ab} &\equiv& {\cal O}(1), \\
 h_{ab} &\equiv& {\cal O}(\epsilon^2), \\
 \nabla_c h_{ab} &=& {\cal O}(\epsilon^2/\kappa)
 \label{eq:derivs} \ .
\end{eqnarray}
\end{mathletters}
Formally, at least, perturbation of a curved background must
parameterize the size of derivatives because the background
radius of curvature is a natural length scale.  In addition,
many cosmological problems have a particle horizon length $L$.
The parameter $\kappa$ represents the length scale of the
perturbations relative to the particle horizon: $\kappa=l/L$.
In principle it can be either greater or less than one; in
practice it is usually small.  Specific restrictions will
apply when we consider orders of magnitude more carefully in
section~\ref{sec:OOM}.

Besides taking $\epsilon^2\ll1$, it is also assumed that
$\epsilon^2/\kappa\ll1$.  As shown below, this implies that the
matter inhomogeneities always move slowly (non-relativistically),
and also that the effective stress-energy of the metric
perturbations is small.

\newpage
\section{The perturbation equations}
The homogeneity and isotropy of the background $\gamma_{ab}$ allows
separation of space and time dependence in the field equations,
so it is useful to write the perturbations not as spacetime
functions $h_{ab}$ but as a harmonic decomposition.  (See, e.g.,
Harrison, 1967~\cite{har67} and Sasaki, 1987~\cite{sas87}.)  The
perturbations' spatial dependence is expanded in eigenfunctions,
or normal modes, of the covariant Laplacian $\nabla_i\nabla^i$ on
the 3D static background $\gamma_{ij}$, reducing the gravitational
field equations to a set of equations for the time-dependent
amplitudes of the modes.  The Appendix gives some general
references to the (extensive) literature on the eigenfunctions.

In this paper we focus on the scalar modes, representing
perturbations to the metric and matter variables that can be
written entirely in terms of solutions to the equations
\begin{mathletters}
\begin{eqnarray}
 \nabla_i \nabla^i Q(\vec{x},\vec{q}) &+& q^2 Q = 0
 \label{eq:Q} \\
 k=+1 &:& q^2 = 0,3,8,\ldots(n^2-1), \\
 k= 0 &:& q^2 \geq 0, \\
 k=-1 &:& q^2 \geq 1\ .
\end{eqnarray}
\end{mathletters}
\label{eq:allQ}
This describes the sort of fields resulting from density
fluctuations.  Spatial derivatives are longitudinal with order of
magnitude $\nabla_i Q = {\cal O}(q) = {\cal O}(\kappa^{-1})$.
Vector (rotational) and tensor (transverse) modes are not
included, since it is reasonable to suppose that most observables
(such as lensing and redshift) are dominated by scalar effects.
However, inflation-induced gravitational waves could have a
significant effect on the anisotropy of the cosmic microwave
background; see, e.g.\ Davis et al., 1992~\cite{davis92}.

Representing the perturbations this way both simplifies and
complicates the interpretation of $\kappa$.  It simplifies, in
that for a given scalar mode $Q$, $\kappa$ is just the reduced
wavelength: $\kappa = q^{-1}$.  It complicates, in that for a real
condensed object like a galaxy or cluster, $\kappa$ must be thought
of as characterizing those wavelengths which contribute the largest
physical effect at a point.  With this in mind, the expansion of
the metric perturbations comes from Bardeen 1980~\cite{bar80}:
\begin{mathletters}
\begin{eqnarray}
 h_{00} &=& 2 \gamma_{00} \int\!\!d\mu(\vec q)\:
            Q(\vec x,\vec q) A(\eta,\vec q), \\
 h_{0i} &=& \int\!\!d\mu(\vec q)\:
            q^{-1} \nabla_i Q(\vec x,\vec q) B(\eta,\vec q), \\
 h_{ij} &=& 2 \gamma_{ij} \int\!\!d\mu(\vec q)\:
            Q(\vec x,\vec q)
            [H(\eta,\vec q) + \frac{1}{3} H_{\rm T}(\eta,\vec q)]
            \nonumber \\
        &&  + 2 \int\!\!d\mu(\vec q)\:
            q^{-2} \nabla_i \nabla_j
            Q(\vec x,\vec q) H_{\rm T}(\eta,\vec q)\ .
\end{eqnarray}
\end{mathletters}
Integrals over the measure $\mu(\vec q)$ stand for whatever
operation expresses the completeness of the $Q$'s, depending on
whether the eigenvalue spectra in equations (\ref{eq:allQ}) give a
discrete or continuous set of modes.  Most linearized calculations
omit the integrals as being implicit, but in this paper we write
them explicitly for clarity in what follows.

Choosing longitudinal gauge, $B=H_{\rm T}=0$, simplifies these
expressions greatly.  With only scalar perturbations this gauge
fully specifies the metric and lets us think of the perturbations
as local length contractions and time dilations, making for easy
comparison with Newtonian theory because it leaves the metric
diagonal:
\begin{mathletters}
\begin{eqnarray}
 h_{00} &=& -2 \int\!\!d\mu\: Q A, \\
 h_{0i} &=& 0, \\
 h_{ij} &=& +2 \gamma_{ij} \int\!\!d\mu\: Q H\ .
\end{eqnarray}
\end{mathletters}
\label{eq:hab}
Note that the amplitudes $A(\eta,\vec q)$ and $H(\eta,\vec q)$
are supposed to have the same order of magnitude ($\epsilon^2$)
as the metric perturbations themselves.

(Other common choices for the gauge include harmonic and
synchronous.  Harmonic gauge is the usual choice for gravitational
waves, and eliminates many higher-order perturbation terms in the
Einstein tensor, but the form of the metric is more complicated.
Synchronous gauge is also common, and has the advantage of
leaving the time coordinate unchanged.  But it suffers from being
under-specified, even when there are only scalar modes, allowing
spurious gauge-mode solutions which can be difficult and annoying
to identify and remove.)

In longitudinal gauge the Einstein tensor is
\begin{mathletters}
\begin{eqnarray}
 G_{00} &=& 3[(\frac{a'}{a})^2 + k] + 2\int\!\!d\mu\: Q
            [q^2 H + 3\frac{a'}{a}H' + 3k(A-H)], \\
 G_{0i} &=& 2\int\!\!d\mu\: \nabla_i Q [\frac{a'}{a}A - H'], \\
 G_{ij} &=& [(\frac{a'}{a})^2 - 2(\frac{a''}{a}) - k] \gamma_{ij}
            \nonumber \\
        &&  - 2\gamma_{ij}\int\!\!d\mu\:
                Q \left\{ \frac{1}{2}q^2(A+H)
                + [(\frac{a'}{a})^2 - 2(\frac{a''}{a})] (A-H)
                - \frac{a'}{a}(A'-2H') + H'' \right\}
            \nonumber \\
        &&  - \int\!\!d\mu\: \nabla_i \nabla_j Q (A+H)\ .
\end{eqnarray}
\end{mathletters}
\label{eq:Gab}
This includes all terms linear in $h_{ab}$ and its derivatives.
It does not include nonlinear terms, which are
${\cal O}(\epsilon^4)$,
${\cal O}(\epsilon^4/\kappa)$,
${\cal O}(\epsilon^4/\kappa^2)$,
or smaller, the so-called pseudotensor terms discussed at the
end of section~\ref{sec:OOM}.  The stress-energy tensor $T_{ab}$
for the matter is constructed by defining variables in the matter
rest frame and then performing a Lorentz boost into the coordinate
frame~\cite{bar80}.  The perfect-fluid background model is given
(scalar) perturbations to density, pressure, and velocity as:
\begin{mathletters}
\begin{eqnarray}
 \tilde{\rho}(\eta,\vec x) &=&
   \rho(\eta) + \rho(\eta) \int\!\! d\mu(\vec q)\:
   Q(\vec x,\vec q) \Delta(\eta,\vec q), \\
 \tilde{P}(\eta,\vec x) &=&
   P(\eta) + \rho(\eta) \int\!\!d\mu(\vec q)\:
   Q(\vec x,\vec q) \Pi(\eta,\vec q), \\
 \tilde{v}_i(\eta,\vec x) &=&
   0 - \int\!\!d\mu(\vec q)\: q^{-1} \nabla_i
   Q(\vec x,\vec q) v(\eta,\vec q)\ .
\end{eqnarray}
\end{mathletters}
Actually, it is better to call $\Delta$ a density
{\em fluctuation\/} because, as mentioned earlier, it is not
necessarily small. However, the changes in velocity and pressure
are always small, as shown below.  (Note that the definition of
the pressure perturbation $\Pi$ differs from Bardeen~\cite{bar80};
this definition is easier to interpret when $P=0$.)

To first order in velocity, the boost to the coordinate frame
gives components of $T_{ab}$ as:
\begin{mathletters}
\begin{eqnarray}
 T_{00} &=& a^2 \rho
            \left[ 1 + \int\!\! d\mu\: Q (2A+\Delta) \right], \\
 T_{0i} &=& a^2 \rho
            \left[ (1+\sigma)\int\!\! d\mu\: q^{-1} \nabla_i Q v
            + \int\!\! d\mu\: q^{-1} \nabla_i Q v
              \int\!\! d\mu\: Q \Delta
            \right], \\
 T_{ij} &=& a^2 \rho \gamma_{ij}
            \left[ \sigma + \int\!\! d\mu\: Q (2 \sigma H + \Pi)
            \right]\ .
\end{eqnarray}
\end{mathletters}
\label{eq:Tab}
($\sigma$ is the background pressure/density ratio.)  Remainders
are ${\cal O}(v^2)$; the product term in the time-space component
is kept because $\Delta$ may be greater than one.

Most approaches to perturbation theory equate terms with equal
orders of magnitude.  But deferring order of magnitude arguments
lets us exploit the harmonic decomposition of the field equations
first, in essentially the same way as isolating coefficients in
a Fourier series [see equation (\ref{eq:intFE})].  This has the
advantage that the results obtained remain valid as the relative
order of magnitude of terms in the field equations change---say
as the density contrast is either diffuse (large $\kappa$) or
condensed (small $\kappa$), or as its amplitude becomes large
or small.

Removing the trace from the Einstein equation
$G_{ij}-8\pi T_{ij}=0$ gives a result which looks
familiar from linearized gravity (on flat spacetimes):
\begin{equation}
 2 \int\!\! d\mu\:
  (\nabla_i \nabla_j Q + \frac{1}{3} q^2 \gamma_{ij} Q)
  (A + H) = 0\ ,
\end{equation}
which can only be true for arbitrary amplitudes $A$ and $H$ if
\begin{equation}
 A(\eta,\vec q) = - H(\eta,\vec q) + {\cal O}(\epsilon^4)\ .
\label{eq:A-H}
\end{equation}
Using this in equations (\ref{eq:hab}) and (\ref{eq:metric}) for
the perturbations and the line element puts the metric in the
linearized pseudo-Newtonian form widely used for studies of
gravitational lensing, etc:
\begin{eqnarray}
 ds^2 &=& a^2 [-(1+2\phi)d\eta^2 + (1-2\phi)\gamma_{ij}dx^i dx^j]\ ,
 \label{eq:metric2}
 \\
 \phi(\eta,\vec x)
      &=& - \frac{1}{2} h_{00}
       =  - \int\!\! d\mu(\vec q)\: Q(\vec x,\vec q) H(\eta,\vec q)
          + {\cal O}(\epsilon^4)\ .
 \label{eq:phi}
\end{eqnarray}
To obtain equations for $H(\eta,\vec q)$, and thus
$\phi(\eta,\vec{x})$, in terms of the matter variables we use
equation (\ref{eq:A-H}) in the components of the orthogonality
equation
\begin{equation}
 \int\!\!dV\: Q^\ast (G_{ab} - 8\pi T_{ab}) = 0 \ ,
\label{eq:intFE}
\end{equation}
where $dV$ is the proper volume element in the static 3-space
$\gamma_{ij}$ (see the Appendix).  The density fluctuations
are governed primarily by the time component $a=b=0$.  Using
equations (\ref{eq:Gab}) and (\ref{eq:Tab}) for the Einstein and
stress-energy tensors gives:
\begin{equation}
 3 (\frac{a'}{a}) H' + (q^2 + 8\pi a^2\rho - 6k) H =
 4\pi a^2\rho\Delta\ .
\label{eq:HDelta}
\end{equation}
To this level of approximation, the scale factor obeys the usual
FRW equation for the background density $\rho(\eta)$ [compare
equation (\ref{eq:a}) below]:
\begin{equation}
 \frac{8\pi}{3} (a^2 \rho) =
 \left( \frac{a'}{a} \right)^2 + k \ .
\end{equation}
In a formal sense, $a$ must obey this equation in order that
the integration in equation (\ref{eq:intFE}) gives a Dirac
delta function in position when applied to the ${\cal O}(1)$
(background) terms in the field equations, owing to the assumed
completeness of the $Q$'s.  This allows us to make free use of
substitutions from the background model for terms involving the
scale factor and $\rho$ in the following sections.

Equations relating the metric perturbations to velocity and
pressure perturbations follow similarly.  The equations for
$\Pi$ and $v$ come most readily from the spatial and space-time
components of (\ref{eq:intFE}) respectively:
\begin{equation}
 H'' + 3 (\frac{a'}{a}) H' - (8\pi a^2 \rho\sigma + 2k) H =
 -4\pi a^2 \rho \Pi \ ,
\label{eq:HPi}
\end{equation}
\begin{equation}
 q [H' + (\frac{a'}{a})H] =
 - 4\pi a^2 \rho (1+\sigma) v [1+{\cal O}(\Delta)] \ .
\label{eq:Hv}
\end{equation}
(Strictly speaking, the last equation holds only for $q\neq0$---no
real restriction, because $q=0$ is a constant mode, merely
representing an improper definition of background.)  We leave the
$\Delta$ correction in (\ref{eq:Hv}) inexplicit since products
of eigenfunctions add nontrivial complications to the formalism,
but this suffices for order of magnitude assessment.

\newpage
\section{Orders of Magnitude}
\label{sec:OOM}
Comparing the perturbations and their relative effects on the
metric requires careful consideration of the size of the time
derivatives.  These are of order one in systems that are not
gravitationally bound, $H'={\cal O}(\epsilon^2)$.  However in bound
systems we expect $H'={\cal O}(\epsilon^3/\kappa)$---for instance,
one can imagine an observer stationed a fixed distance $R$ from
the center of a collapsing dust cloud; changes in the potential
are $\sim \phi(V/R)$, where $V$ is the speed of infall for the
dust, or $\sim \epsilon^2 (\epsilon/\kappa)$ by order of magnitude.

In equation (\ref{eq:HDelta}), the size of the density contrast
depends on the ratio $\epsilon/\kappa$ [remember that
$q={\cal O}(\kappa^{-1})$].  The allowed regimes are labeled as
linear or nonlinear density, depending on the size of $\Delta$:
\begin{quasitable}
\begin{tabular}{rc|cc}
 {}
    & ${\cal O}(H')$
    & ${\cal O}(\Delta)$,\ $\kappa\ll1$
    & ${\cal O}(\Delta)$,\ $\kappa\gg1$
 \\ \hline
 LDR: $\epsilon/\kappa\ll1$
    & $\epsilon^2$
    & $\epsilon^2/\kappa^2$
    & $\epsilon^2$
 \\
 NLDR: $\epsilon/\kappa\gg1$
    & $\epsilon^3/\kappa$
    & $\epsilon^2/\kappa^2$
    & not allowed ($\epsilon\gg1$)
 \\
 {} & {} & {}
\end{tabular}
\end{quasitable}
Strong density fluctuations with large (super-horizon) scales are
not allowed because they create potentials that cannot be treated
as perturbations on the background metric.  (So the results derived
here, in particular the Green function given later, should not be
expected to work in a strongly ``tilted'' universe.)  However, the
small-scale density fluctuations (NLDR with $\kappa\ll1$) should
have an order of magnitude consistent with Newtonian theory.
The prediction from the Poisson equation
\begin{equation}
 \nabla^2 \phi =
 -\frac{1}{2a^2} \nabla_i\nabla^i h_{00} =
 4\pi \rho \Delta
\label{eq:Poisson}
\end{equation}
is also that $\Delta$ should be ${\cal O}(\epsilon^2/\kappa^2)$.
This holds for weak density fluctuations as well, again provided
that the scale is small, since Newtonian physics can be expected
to apply in a sufficiently small region of space.

Orders of magnitude for pressure and velocity perturbations come
{}from equations (\ref{eq:HPi}) and (\ref{eq:Hv}):
\begin{quasitable}
\begin{tabular}{r|cc}
 {}
    & ${\cal O}(\Pi)$
    & ${\cal O}(v)$
 \\ \hline
 LDR: $\epsilon/\kappa\ll1$
    & $\epsilon^2$
    & $\epsilon^2/\kappa$
 \\
 NLDR: $\epsilon/\kappa\gg1$
    & $\epsilon^4/\kappa^2$
    & $\epsilon$
 \\
 {} & {} & {}
\end{tabular}
\end{quasitable}
These agree nicely with the simple Newtonian argument that for
bound systems (NLDR) the velocity should be proportional to the
square root of the potential,
$v \sim H^{1/2} \sim \epsilon$,
and the pressure should be
$\Pi \sim \Delta v^2 \sim \epsilon^4/\kappa^2$.
More generally we can use the Euler equation for the hydrodynamics
of a perfect fluid in a gravitational field:
\begin{equation}
 \frac{d\tilde{\vec{v}}}{dt} =
 \left[ \partial_t\tilde{\vec{v}} +
 (\tilde{\vec{v}}\cdot\vec{\nabla}) \tilde{\vec{v}} \right] =
 -\tilde{\rho}^{-1} \vec{\nabla}\tilde{P} - \vec{\nabla}\phi \ .
\label{eq:Euler}
\end{equation}
In the linear density regime the partial time derivative is
most important, and comparing orders of magnitude we find
$v={\cal O}(\epsilon^2/\kappa)$ and $\Pi={\cal O}(\epsilon^2)$,
as in the table above.  In the non-linear density regime
$\tilde{\rho}$ is of order $\epsilon^2/\kappa^2$, and again the
estimates agree with the results from equations (\ref{eq:HPi}) and
(\ref{eq:Hv}).

The tables show, not surprisingly, that in any allowed (subhorizon)
regime, the pressure and velocity perturbations are much weaker
than the density fluctuations.  So under these conditions the
metric perturbations $H(\eta,\vec q)$ are determined primarily
by $\Delta(\eta,\vec q)$; that is, hydrodynamically the density
fluctuations can be treated as the source.

We also have to consider effects on the scale factor $a$, since it
makes an implicit contribution to any order of magnitude arguments.
When nonlinear terms are kept in the Einstein tensor, their
spatial average can be thought of as an energy density and used
to construct an effective stress-energy tensor for the metric
perturbations.  (See, e.g., Isaacson, 1968b~\cite{isac68b}.)
Dropping for a moment the requirement that $a(\eta)$ come from
the background, on physical grounds we expect
\begin{equation}
 a(\eta) = a_{\rm FRW} [1 + {\cal O}(<\epsilon^4/\kappa^2>)]
\label{eq:a}
\end{equation}
but clearly, the average over even a relatively small volume must
be less than the maximum value; i.e.
${\cal O}(<\epsilon^4/\kappa^2>) \ll \epsilon^4/\kappa^2 \ll 1$.
So using the background scale factor does not alter the arguments
about $\Delta$, $\Pi$, and $v$, simply because the corrections are
even smaller than any of the terms discussed previously.

\newpage
\section{Green Function}
Having determined that the scalar metric perturbations are
determined primarily by the density fluctuations, we can relate
them directly by solving equation (\ref{eq:HDelta}) for the mode
amplitudes $H(\eta,\vec{q})$ and using the result in (\ref{eq:phi})
to get $\phi(\eta,\vec{x})$.  Equation (\ref{eq:HDelta}) is of
the Riccati type; the standard solution given in handbooks like
Gradshteyn and Ryzhik~\cite{gr} is
\begin{equation}
 H(\eta,\vec q) =
 H(\eta_0,\vec q) E(\eta_0,\eta,q) -
 \int_{\eta_0}^{\eta}\!\!du\: I(u,\vec q) E(u,\eta,q)
\end{equation}
where
\begin{mathletters}
\begin{eqnarray}
 E(u,\eta,q) &=& \frac{a(u)}{a(\eta)}
                 \exp[-(q^2-3k)C(u,\eta)], \\
 I(u,\vec q) &=& -\frac{4\pi}{3}
                 \left( \frac{a^3\rho}{a'} \right)_u
                 \Delta(u,\vec q), \\
 C(u,\eta) &=& \frac{1}{3}
                 \int_{u}^{\eta}\!\!dv\: (\frac{a}{a'}) \ .
\end{eqnarray}
\end{mathletters}
All of this, along with the definitions
\begin{mathletters}
\begin{eqnarray}
 H(\eta_0,\vec q) &=& -\int\!\!dV(\vec{y})\:
   Q^\ast(\vec y,\vec q) \phi(\eta_0,\vec y), \\
 \Delta(u,\vec q) &=& \int\!\!dV(\vec{y})\:
   Q^\ast(\vec y,\vec q) \Delta(u,\vec y),
\end{eqnarray}
\end{mathletters}
is inserted into equation (\ref{eq:phi}).  The integrals are
re-ordered to contract the mode sums as much as possible, to give
a kernel against which the initial conditions
$\phi(\eta_0,\vec y)$ and source $\Delta(u,\vec y)$
are integrated over space.  After some manipulation this gives:
\begin{eqnarray}
 \phi(\eta,\vec x) &=& \int\!\!dV(\vec{y})\:
      G(\eta_0,\eta,\vec x,\vec y) \phi(\eta_0,\vec y)
 \nonumber \\
 & & -\frac{4\pi}{3} \int_{\eta_0}^{\eta}\!\! du\:
      \frac{a^3\rho}{a'} \int\!\!dV(\vec{y})\:
      G(u,\eta,\vec x,\vec y) \Delta(u,\vec y)
      + {\cal O}(\epsilon^4)
\label{eq:phiG}
\end{eqnarray}
where
\begin{equation}
 G(u,\eta,\vec x,\vec y) =
 \frac{a(u)}{a(\eta)} e^{3kC(u,\eta)}
 \int\!\! d\mu(\vec q)\: Q(\vec x,\vec q) Q^\ast(\vec y,\vec q)
      e^{-q^2 C(u,\eta)} \ ;
\label{eq:genG}
\end{equation}
which is the central result of this paper: a Green function
for metric perturbations due to scalar density fluctuations in a
Robertson-Walker background.  Equation (\ref{eq:phiG}) has several
properties expected from an expression for a pseudo-Newtonian
potential; for instance an overdensity $\Delta(u,\vec y) > 0$
decreases $\phi$ in its neighborhood.  Other simple cases can
be checked with the help of the following formulas for special
arguments of the Green function, which come from the completeness
and orthogonality relations, respectively:
\begin{eqnarray}
  G(u,u,\vec x,\vec y) &=& \delta(\vec{x},\vec{y}),
  \label{eq:Gdelta}
  \\
  \int\!\!dV\: G(u,\eta,\vec x,\vec y) &=&
    \frac{a(u)}{a(\eta)} e^{3kC(u,\eta)} \ .
  \label{eq:specialG}
\end{eqnarray}
The first shows that $\phi$ matches its initial conditions
as $\eta\rightarrow\eta_0$ in (\ref{eq:phiG}).  The second,
which holds as long as $Q={\rm const}$ is an allowed mode,
shows that $\phi$ is a function of time alone if $\Delta$ is,
which means only that the background density has been shifted:
$\rho\rightarrow\rho\Delta(\eta)$; of course $\Delta=0$ is also
a solution.  (Any spatially homogeneous form for the density
fluctuation leads to a standard Robertson-Walker metric after
transformation of the scale factor and time coordinate.)

Specific forms for the Green function (\ref{eq:genG}) come from
choosing a particular representation (a coordinate system, in other
words) for the $Q$'s, given $k$.  Regardless of the representation,
in closed or open spaces ($k=\pm1$) the integrals can be very ugly,
especially for $k=-1$.  But for angles smaller than the curvature
scale $k=0$ is a good approximation, and under inflation it would
hold generally.  In this case we can represent the $Q$'s as plane
waves and carry out the integrals explicitly.  Using rectangular
coordinates for $\vec{q}$, the integral in (\ref{eq:genG}) is
separable, and completing the square in each exponent gives
\begin{equation}
 G_{k=0}(u,\eta,\vec x,\vec y) =
 \frac{a(u)}{a(\eta)}
 \frac{1}{[4\pi C(u,\eta)]^{3/2}}
 \exp\left[-\frac{|\vec y - \vec x|^2}{4C(u,\eta)}\right] \ .
\label{eq:G0}
\end{equation}
The rest of the paper is devoted primarily to examining this
formula and its use in equation (\ref{eq:phiG}) for the potential.
Henceforth $k$ is implicitly zero, unless noted otherwise.

\newpage
\section{Analogy with Diffusion}
The most striking thing about equation (\ref{eq:G0}) is that
it looks very much like the Green function for diffusion in a
uniform medium, and more generally, the displacement probability
distribution for an isotropic random walk \cite{chan43}.
For instance, we can compare the solution to the equation of
heat conduction:
\begin{equation}
 \frac{\partial T}{\partial t} =
 \kappa\nabla^2 T + c_{\rm p}^{-1} \dot{q} \ .
\label{eq:heat}
\end{equation}
(In this section, $\kappa$ and $\dot{q}$ represent thermal
diffusivity and the rate of heat production per unit mass, not the
reduced wavelength and mode numbers used in previous sections.)
In an infinite uniform medium with no sources ($\dot{q}=0$)
the solution for an arbitrary initial temperature distribution
is \cite{fw}
\begin{mathletters}
\begin{eqnarray}
 T(\vec{r},t) &=&
   \int\!\! d^3\!\vec{r}\,'\:
   G(\vec{r}-\vec{r}\,',t-t_0)
   T(\vec{r}\,',t_0),
\\
 G(\vec{r}-\vec{r}\,',t-t_0) &=&
   \frac{1}{[4\pi\kappa(t-t_0)]^{3/2}}
   \exp\left[
   -\frac{|\vec{r}-\vec{r}\,'|^2}{4\kappa(t-t_0)}
   \right] \ .
 \label{eq:temp}
\end{eqnarray}
\end{mathletters}
Since there are no sources, the gravitational analog is the
integral over initial conditions in equation (\ref{eq:phiG}) for
the potential, using the Green function (\ref{eq:G0}).  If the
integral for the potential is written with a physical volume
element $a^3(\eta_0)dV$, like the temperature integral above,
we compare $a^{-3}(\eta_0) G(\eta_0,\eta,\vec{x},\vec{y})$ with
equation (\ref{eq:temp}).  Taking $\eta_0$ and $\eta$ close
together to minimize the kinematic effects of expansion, the
analog of the diffusivity and time factors is just
\begin{equation}
 \kappa(t-t_0) \rightarrow a^2(\eta_0) C(\eta_0,\eta)
               \simeq (3H_0)^{-1} \Delta t
\label{eq:difftime}
\end{equation}
where $H_0$ is the Hubble parameter, evaluated at $\eta_0$ in
this particular instance, and $\Delta t$ is the proper (rather
than conformal) time interval corresponding to $(\eta-\eta_0)$.
This analog of the diffusivity is consistent with an alternate
approach that comes from converting equation (\ref{eq:HDelta})
for the mode amplitudes $H(\eta,\vec{q})$ to an equation
in $(\eta,\vec{x})$, to form the analog of (\ref{eq:heat}).
Summed over modes and converted to physical variables (proper
time, standard Laplacian, etc) it can be written as
\begin{equation}
 \partial_t (a\phi) =
 (3H_0)^{-1} \nabla^2 (a\phi) -
 \frac{1}{2} H_0 (a\Delta) \ ,
\label{eq:phidiffusion}
\end{equation}
where the Hubble parameter is taken to be a function of time.
This is a diffusion equation for the potential, with time-dependent
diffusivity and ``specific heat capacity:''
\begin{mathletters}
\begin{eqnarray}
 T         &\leftrightarrow&  -a\phi      \ , \\
 \kappa    &\leftrightarrow&  (3H_0)^{-1} \ , \\
 c_{\rm p} &\leftrightarrow&  2H_0^{-1}   \ , \\
 \dot{q}   &\leftrightarrow&  a\Delta     \ ,
\end{eqnarray}
\end{mathletters}
in agreement with (\ref{eq:difftime}).  In simple kinetic theory,
the diffusivity would imply a mean free path of $H_0^{-1}$
for particles of average speed $c=1$.  In this approximation,
changes in $\phi$ travel at all speeds, as can be seen both
{}from the analogy with diffusion and from the fact that equation
(\ref{eq:phiG}) assigns a non-zero value to the potential
everywhere, even for (spatially) localized density fluctuations.
This superficial causality problem comes from dropping terms
${\cal O}(\epsilon^4/\kappa^2)$ while constructing
(\ref{eq:HDelta}).  Ordinary diffusion cures the causality
problem by making the diffusivity and specific heat capacity
temperature-dependent \cite{hd}.  In the gravitational case
this means making $H_0=a'a^{-2}$ a function of the metric
perturbations---which it is, in an exact treatment, since the scale
factor is affected by the spatial average of $(\vec{\nabla}\phi)^2$
(cf equation (\ref{eq:a})).  But causality does {\em not\/}
represent a problem in the application of (\ref{eq:phiG}), however,
because far from the source, ``errors'' in $\phi$ are extremely
small---below the level of approximation for the calculation.

\newpage
\section{Newtonian Limit}
The diffusion analog is easiest to see when considering the role of
the initial conditions.  The Newtonian limit treats the opposite
situation, where we consider compact sources (gravitationally
bound systems, for instance) evolving slowly and with negligible
initial conditions.  If the time derivative is ignored, the
gravitational diffusion equation (\ref{eq:phidiffusion}) becomes
the Poisson equation (\ref{eq:Poisson}) for the potential, after
replacing $H_0^2$ by $8\pi G \rho /3$.  This makes it reasonable
to suppose that the Green-function expression (\ref{eq:phiG}) for
the potential, which is essentially derived from the gravitational
diffusion equation, can be reduced to a Newtonian form under
appropriate conditions, providing a useful check of the formula.

We use the formulas for $k=0$, but in fact the Newtonian
limit holds in a sufficiently small region of any cosmological
metric~\cite{peebles}, and so the results can be expected to
hold for $k=\pm 1$ as well.  (Even when $k\neq 0$ the diffusion
equation (\ref{eq:phidiffusion}) still reduces to the Poisson
equation when only the NLDR leading terms are kept.)  Using the
background equations for the scale factor and density we write
the potential as:
\begin{equation}
 \phi = -\int\!\!dV\:
        \left\{ \frac{1}{2}\int_{\eta_0}^{\eta}\!\! du\:
        \frac{a'}{a} G \Delta - G_0 \phi_0 \right\}
\end{equation}
by analogy with the Newtonian form.  The contribution of the
initial conditions $G_0\phi_0$ is negligible provided that
$\eta_0$ is taken far in the past and we assume that the early
universe was smooth.  Then to evaluate the time integral, we can
imagine that the potential is to be measured at a point outside
a nearby ``lump'' of matter.  In this situation it is physically
reasonable to suppose that the Green function will peak for values
of conformal time $u$ close to $\eta$, corresponding to the time
it takes for changes in the source to influence the observer.
For small values of reversed conformal time $w=\eta-u$ the Green
function is
\begin{mathletters}
\begin{eqnarray}
 G &=& \left( \frac{3}{4\pi} \frac{a'}{a} \right)^{3/2}
       w^{-3/2} \exp(-D_0/w) + {\cal O}(w^{-1/2}), \\
 D_0 &\equiv& \frac{3a'}{a} \frac{|\vec{y}-\vec{x}|^2}{4} \ .
\end{eqnarray}
\end{mathletters}
This shows a peak at approximately $w_{\rm p}=(2/3)D_0$,
with a width of about $(2/5)D_0$.  Both are small provided
that the matter distribution $\Delta(\eta,\vec{y})$ is such
that the volume integral restricts $D_0$ to small values.
More precisely, an expansion in $w$ is self-consistent if the
width of the peak is small compared to the elapsed conformal
time: $D_0\ll(\eta-\eta_0)$.  With the flat background, and using
$\eta_0=0$, this means that (omitting overall constants of order
one) we must have
\begin{mathletters}
\begin{eqnarray}
 \frac{|\vec{y}-\vec{x}|^2}{\eta^2} &\ll& 1,\ \ {\rm or} \\
 \kappa^2 &\ll& 1 \ ,
\end{eqnarray}
\end{mathletters}
where the last line follows from the ``original'' definition of
$\kappa$ as perturbation size divided by particle horizon length.
Thus the Newtonian limit holds for density perturbations that are
highly localized, as one would expect.  Furthermore, if the density
field changes little in the time corresponding to the width of
the peak, the source and expansion terms can be evaluated at $w=0$
and moved outside the time integral.  (Strictly speaking,
$w_{\rm p}$ might be a better choice, amounting to a notion of
``retarded time,'' but the errors due to approximating the Green
function make this academic.)  Another change of variable to
$z=D_0/w$ puts the time integral in nearly standard form, and
we find:
\begin{mathletters}
\begin{eqnarray}
 \phi &\simeq& -\int\!\! dV\:
    \left\{ \frac{1}{2} \left(\frac{3}{4\pi}\right)^{3/2}
      \left(\frac{a'}{a}\right)^{5/2} \frac{\Delta}{D_0^{1/2}}
      \int_{\delta}^{\infty}\!\!dz\:z^{-1/2} e^{-z}
    \right\},
 \label{eq:phierfc}
 \\
 \delta &\equiv& \frac{D_0}{\eta-\eta_0}
    = \frac{3}{4} H_0 \frac{a^2|\vec{y}-\vec{x}|^2}{a(\eta-\eta_0)}
    \ll 1 \ .
 \label{eq:D0}
\end{eqnarray}
\end{mathletters}
The $z$-integral is simply $\pi^{1/2}{\rm erfc}(\delta^{1/2})$, or,
to a good approximation when $\delta$ is small, just $\pi^{1/2}$.
This along with the definition of $D_0$ and the ever-present
background equations for the scale factor reduce (\ref{eq:phierfc})
to the familiar Newtonian form:
\begin{eqnarray}
 \phi &\simeq& -\int\!\! dV\:
    \left\{ \frac{3}{8\pi} \left( \frac{a'}{a} \right)^2
      \frac{\Delta}{|\vec{y}-\vec{x}|}
    \right\} \nonumber \\
 &\simeq& -\int\!\! a^3 dV\:
    \frac{\rho\Delta}{a|\vec{y}-\vec{x}|} \ .
\end{eqnarray}
To summarize, this formula holds when the initial conditions can
be neglected, and the spatial dependence of the density contrast
$\Delta(u,\vec{y})$ limits $\delta$ to small values in equation
(\ref{eq:phierfc}), while the time dependence has a scale much
larger than (any value of) $D_0$.  In fact most gravitationally
bound systems satisfy these criteria quite well, and are also quite
uninteresting from a cosmological standpoint.  More interesting are
situations where the time evolution of the density fluctuations
makes a significant contribution to the metric.  A theoretical
description requires extending the calculation just given to
post-Newtonian order by expanding the Green function and source
terms in a time-series.  We hope to show in a forthcoming paper
\cite{jw93} how this is done, and under what conditions equation
(\ref{eq:phiG}) for the potential predicts significant deviations
{}from the Newtonian (and LDR) approximations.

\newpage
\section{conclusion}
In most situations of observational interest, the Green function
expression (\ref{eq:phiG}) for the pseudo-Newtonian potential
offers a simple and relativistically correct way of calculating
the metric perturbations, taking into account effects such as
multiple (perhaps closely spaced) sources, deviations from the
``thin lens'' approximation, non-linear density evolution, and
the cosmological expansion.  The results can be applied in the
calculation of observational effects such as lensing, redshift,
and time-delay.  We hope that this will stimulate exploration of
situations that are difficult to treat with current techniques.
A upcoming paper will examine the post-Newtonian limit of the
Green-function expression, with attention to those situations
which predict significant observable effects.

\newpage
\acknowledgments
This work was supported in part by NASA grant NAS8-36125
at Stanford University and NAGW-763 at Steward Observatory.

MWJ would like to thank Dr.\ Joseph Keller of the Stanford
University Mathematics Department for his comments, and
Dr.\ Glenn Ierely of the UCSD Institute of Geophysics and Planetary
Physics for pointing out the ``cure'' to the causality problem
in diffusion.  He also thanks Peter Schneider of the Max Planck
Institute and Lee Lindblom of Montana State University for
helpful discussions.

Both MWJ and EVL acknowledge the generous support of the Gravity
Probe-B group and the Center for Space Science and Astrophysics at
Stanford University.

\newpage
\unletteredappendix{}
This appendix gives selected references to the literature
on harmonic functions.  Most calculations do not require an
explicit representation; in spherical coordinates a schematic
representation is:
\begin{eqnarray*}
 Q(\vec{x},\vec{q}) &=&
   \Pi_{l}^{(k)}(n,\alpha) Y_{lm}(\theta,\phi), \\
 \vec{x} &=& (\alpha,\theta,\phi), \\
 \vec{q} &=& (n,l,m), \\
 q^2 &=& n^2 - k,
\end{eqnarray*}
with the eigenvalue spectra in equations (\ref{eq:allQ}).
The angular functions are ordinary spherical harmonics; good
overviews of the properties of the radial functions and/or
the $Q$'s as a whole are in
Harrison 1967~\cite{har67},
Bardeen 1980~\cite{bar80},
Kodama and Sasaki 1984~\cite{ks84},
and Birrell and Davies 1989~\cite{bd89}.
The last writes the orthogonality and completeness relations as:
\begin{eqnarray*}
 \int\!\! dV\: Q^\ast(\vec{y},\vec{q}) Q(\vec{y},\vec{p}) &=&
 \delta(\vec{q},\vec{p}) \ , \\
 \int\!\! d\mu\: Q^\ast(\vec{x},\vec{q}) Q(\vec{y},\vec{q}) &=&
 \delta(\vec{x}-\vec{y}) \ ,
\end{eqnarray*}
where $dV \equiv (\gamma_i^i)^{1/2} d^3\!y$ is the proper volume
element in the static background 3-space and $d\mu$ is the
measure associated with the eigenvalue spectrum.  More detailed
information, including explicit representations of the radial
eigenfunctions and proofs of orthogonality and completeness
can be found in (see also Harrison, above):
Parker and Fulling 1974~\cite{pf74},
and Abbott and Schaeffer 1986~\cite{as86}.

Finally, note that the literature aimed at problems in quantum
field theory uses only scalar harmonics, while in general
relativity an arbitrary tensor function may be composed of scalar,
vector, and tensor harmonics.  The early sections of
Kodama and Sasaki~\cite{ks84}
give a good explanation of the distinction.

\end{document}